\documentclass{elsart}

\usepackage{amsmath,amssymb,amscd,graphicx,subfig}

\newtheorem{theorem}{Theorem}

\newtheorem{proposition}{Proposition}
\newtheorem{definition}{Definition}

\newcommand{\D}{{\mathrm{d}}}

\begin{document}
\small{Physica A 390 (2011) 1009--1025}

\begin{frontmatter}

\title{Kinetic Path Summation, \\ Multi--Sheeted Extension of Master Equation, \\ and Evaluation of Ergodicity Coefficient}
\author{A.~N.~Gorban \corauthref{cor1}}
 \ead{ag153@le.ac.uk}
\address{University of Leicester, UK}
\corauth[cor1]{Corresponding author: University of Leicester, LE1
7RH, UK}

\maketitle

\begin{abstract}
We study the Master equation with time--dependent coefficients,
a linear kinetic equation for the Markov chains or for the
monomolecular chemical kinetics. For the solution of this
equation a path summation formula is proved. This formula
represents the solution as a sum of solutions for simple
kinetic schemes (kinetic paths), which are available in
explicit analytical form. The relaxation rate is studied and a
family of estimates for the relaxation time and the ergodicity
coefficient is developed. To calculate the estimates we
introduce the {\em multi--sheeted extensions} of the initial
kinetics. This approach allows us to exploit the internal
(``micro")structure of the extended kinetics without
perturbation of the base kinetics.
\end{abstract}
\end{frontmatter}

\section{Introduction \label{sec1}}

\subsection{The problem}
First-order kinetics form the simplest and well-studied class of
kinetic systems. It includes the continuous-time Markov chains
\cite{MeynNets2007,MeynMarkCh2009} (the Master Equation
\cite{VanKampen}), kinetics of monomolecular and pseudomonomolecular
reactions \cite{LumpWei2}, provides a natural language for
description of fluxes in networks and has many other applications,
from physics and chemistry to biology, engineering, sociology, and
even political science.

At the same time, the first-order kinetics are very fundamental and
provide the background for kinetic description of most of nonlinear
systems: we almost always start from the Master Equation (it may be
very high-dimensional) and then reduce the description to a lower
level but with nonlinear kinetics.

For the description of the first order kinetics we select the
species--concentration language of chemical kinetics, which is
completely equivalent to the states--probabilities language of the
Markov chains theory and is a bit more flexible in the normalization
choice: the sum of concentration could be any positive number, while
for the Markov chains we have to introduce special ``incomplete
states".

The first-order kinetic system is weakly ergodic if it allows the
only conservation law: the sum of concentration.   Such a system
forgets its initial condition: the distance between any two
trajectories with the same value of the conservation law tends to
zero when time goes to infinity. Among all possible distances, the
$l_1$ distance ($\| x \|_{l_1}=\sum_i |x_i|$) plays a special role:
it decreases monotonically in time for any first order kinetic
system. Further in this paper, we use the $l_1$ norm on the space of
concentrations.

Straightforward analysis of the relaxation rate for a linear system
includes computation of the spectrum of the operator of the shift in
time. For an autonomous system, we have to find the ``slowest"
nonzero eigenvalue of the kinetic (generator) matrix. For a system
with time--dependent coefficients, we have to solve the linear
differential equations for the fundamental operator (the shift in
time). After that, we have to analyze the spectrum of this operator.
Beyond the simplest particular cases there exist no analytical
formulas for such calculations.

Nevertheless, there exists the method for evaluation of the
contraction rate for the first order kinetics, based on the
analysis of transition graph. For this evaluation, we need to
solve kinetic equations for some irreversible acyclic
subsystems which we call the {\em kinetic paths}
(\ref{KinPathSch}). These kinetic paths are combined from
simple fragments of the initial kinetic systems. For such
systems, it is trivial to solve the kinetic equations in
quadratures even if the coefficients are time--dependent. The
explicit recurrent formulas for these solutions are given
(\ref{reccurentPathSolution}).

We construct the explicit formula for the solution of the
kinetic equation for an arbitrary system with time--dependent
coefficients by the summation of solutions of an infinite
number of kinetic paths (\ref{summationConcentr}).

On the basis of this  summation formula we produce a
representation of the $l_1$ contraction rate for weakly ergodic
systems (\ref{l1Deriv}). Because of monotonicity, any partial
sum of this formula gives an estimate for this contraction.

To calculate the estimates we introduce the {\em multi--sheeted
extensions} of the initial kinetics. Such a multi--sheeted
extension is a larger Markov chain together with a projection
of its (the larger) state space on the initial state space and
the following property: the projection of the extended random
walk is a random walk for the initial chain
(Section~\ref{Sec:MultiSheeted}).

This approach allows us to exploit the internal
(``micro")structure of the extended kinetics without
perturbation of the base kinetics.

It is difficult to find, who invented the kinetic path
approach. We have used it in 1980s \cite{Ocherki}, but consider
this idea as a scientific ``folklore".

In this paper we study the backgrounds of the kinetic path
methods. This return to backgrounds is inspired, in particular,
by the series of work \cite{Sun2006,Sun2007}, where the kinetic
path summation formula was introduced (independently, on
another material and with different argumentation) and applied
to analysis of large stochastic systems. The method was
compared to the kinetic Gillespie algorithm
\cite{Gillespie1977} and on model systems it was demonstrated
\cite{Sun2007} that for ensembles of rare trajectories far from
equilibrium, the path sampling method performs better.

For the linear chains of reversible semi-Markovian processes
with nearest neighbors hopping, the path summation formula was
developed with counting all possible trajectories in Laplace
space \cite{Flomenbom2005}. Higher order propagators and the
first passage time were also evaluated. This problem statement
was inspired, in particular, by the evolving field of single
molecules (for more detail see \cite{Flomenbom2007}).

The idea of kinetic path with selection of the dominant paths
gives an effective generalization of the limiting step
approximation in chemical kinetics
\cite{GorbaRadul2008,GorbanRadZinChemEngSci2010}.

\section{Basic Notions \label{secBasNot}}

Let us recall the basic facts about  the first-order kinetics. We
consider a general network of linear reactions. This network is
represented as a directed graph (digraph)
(\cite{Yab1991,Temkin1996}): vertices correspond to components $A_i$
($i=1,2,\ldots,n$, edges correspond to reactions $A_i \to A_j$
($i\neq j$). For the set of vertices we use notation $\mathcal{A}$,
and for the set of edges notation $\mathcal{E}$. For each vertex,
$A_i \in \mathcal{A}$, a positive real variable $c_i$
(concentration) is defined. Each reaction $A_i \to A_j$ is
represented by a pair of numbers $(i,j)$, $i\neq j$. For each
reaction $A_i \to A_j$ a nonnegative continuous bounded function,
the reaction rate coefficient (the variable ``rate constant")
$k_{ji}(t)\geq 0$ is given. To follow the standard notation of the
matrix multiplication, the order of indexes in $k_{ji}$ is always
inverse with respect to reaction: it is $k_{j\leftarrow i}$, where
the arrow shows the direction of the reaction. The kinetic equations
have the form
\begin{equation}\label{kinurT}
\frac{\D c_i}{\D t}=\sum_{j,\ j\neq i} (k_{ij}(t) c_j -
k_{ji}(t)c_i),
\end{equation}
or in the vector form:  $\dot{c} = K(t)c$. The quantities $c_i$
are concentrations of $A_i$ and $c$ is a vector of
concentrations. We don't assume any special relation between
constants, and consider them as independent quantities.

For each $t$, the matrix of kinetic coefficients $K$ has  the
following properties:
\begin{itemize}
\item{non-diagonal elements of $K$ are non-negative;}
\item{diagonal elements of $K$ are non-positive;}
\item{elements in each column of $K$ have zero sum.}
\end{itemize}
This family of matrices coincides with the family of generators
of finite Markov chains in continuous time
(\cite{MeynNets2007,MeynMarkCh2009}).

A {\it linear conservation law} is a linear function defined on
the concentrations $b(c)= \sum_{i} b_i c_i$, whose value is
preserved by the dynamics \eqref{kinurT}. Equation
\eqref{kinurT} always has a linear conservation law:
$b^0(c)=\sum_i c_i = {\rm const}$.

Another important and simple property of this equation is the
preservation of positivity for the solution of \eqref{kinurT}
$c(t)$: if $c_i(t_0)\geq 0$ for all $i$ then $c_i(t_1)\geq 0$ for
$t_1>t_0$.

For many technical reasons it is useful to discuss not only
positive solutions to \eqref{kinurT} and further we do not
automatically assume that $c_i \geq 0$.

The time shift operator which transforms $c(t_0)$ into $c(t)$
is $U(t,t_0)$. This is a column-stochastic matrix:
$$u_{ij}(t,t_0) \geq 0 \ , \ \ \sum_i u_{ij}(t,t_0) =1 \ \ (t\geq
t_0) \ .$$ This matrix satisfies the equation:

\begin{equation}\label{kinurTU}
\frac{\D U(t,t_0)}{\D t}=KU(t,t_0) \  \mbox{ or } \ \frac{\D u_{il}}{\D t}=  \sum_j (k_{ij}(t) u_{jl} - k_{ji}(t)u_{il})
\end{equation}
with initial conditions $U(t_0,t_0)=\mathbf{1}$, where
$\mathbf{1}$ is the unit operator
($u_{ij}(t_0,t_0)=\delta_{ij}$).

Every stochastic in column operator $U$ is a contraction in the
$l_1$ norm on the invariant hyperplanes $\sum_i c_i = const$.
It is sufficient to study the restriction of $U$ on the
invariant subspace $\{x\ | \ \sum_i x_i=0\}$:
 $$\|U x\| \leq \delta \|x\| \;
 {\rm if } \; \sum_i x_i=0$$
for some $\delta \leq 1$. The minimum of such $\delta$ is
$\delta_U$, the norm of the operator $U$ restricted to its invariant
subspace $\{x\ | \ \sum_i x_i=0\}$. One of the definitions of {\em
weak ergodicity} is $\delta < 1$ \cite{DobrushinErgCoeff1956}. The
unit ball of the $l_1$ norm restricted to the subspace $\{x\ | \
\sum_i x_i=0\}$ is a polyhedron with vertices
\begin{equation}\label{ballVert}
g^{ij}=\frac{1}{2}(e^i-e^j), \;\; i \neq j\ ,
\end{equation}
where $e^i$ are the standard basis vectors in $\mathbb{R}^n$:
$e^i_k=\delta _{ik}$, $\delta _{ik}$ is the Kronecker delta.
For a norm with the polyhedral unit ball, the norm of the
operator $U$ is $$\max_{v\in V}\|U(v)\| \ ,$$ where $V$ is the
set of vertices of the unit ball. Therefore, for a ball with
vertices (\ref{ballVert})
\begin{equation}\label{normSubSp}
\delta_U=\|U\|=\frac{1}{2}\max_{i,j}\sum_k |u_{ki}-u_{kj}|\leq 1 \ .
\end{equation}
This is a half of the maximum of the $l_1$ distances between
columns of $U$. The {\em ergodicity coefficient},
$\varepsilon_U=1-\delta_U$, is zero for a matrix with unit norm
$\delta_U=1$ and one if $U$ transforms any two vectors with the
same sum of coordinates in one vector ($\delta_U=0$).

The contraction coefficient $\delta_U$ (\ref{normSubSp}) is a
norm of operator and therefore has a ``submultiplicative"
property: for two stochastic in column operators $U,W$ the
coefficient $\delta_{UW}$ could be estimated through a product
of the coefficients
\begin{equation}\label{submulti}
\delta_{UW} \leq \delta_{U} \delta_{W}\ .
\end{equation}
We will systematically use this property in such a way. In many
estimates we find an upper border $1 \geq \delta(\tau)\geq
\delta_{U(t_1+\tau,t_1)}$, $t_2 \geq t_1$. In such a case,
$\delta_{U(t_1+\tau,t_1)} \to 0$ exponentially with $\tau \to
\infty$. Nevertheless, the estimate $\delta(\tau)$ may originally
have a positive limit $\delta(\tau) \to \delta_{\infty}>0$ when
$\tau \to \infty$. In this situation we can use $\delta(\tau)$ for
bounded $\tau < \tau_1$ and for $\tau >\tau_1$ exploit the
multiplicative estimate (\ref{submulti}). The moment $\tau_1$ may be
defined, for example, by maximization of the negative Lyapunov
exponent:
\begin{equation}\label{expmax}
\tau_1={\rm arg}\max_{\tau>0}\left\{-\frac{\ln(\delta(\tau))}{\tau}\right\} \ .
\end{equation}

For a system with external fluxes $\Pi_i(t)$ the kinetic
equation has the form

\begin{equation}\label{kinurTflux}
\frac{\D c_i}{\D t}=\sum_j (k_{ij}(t) c_j - k_{ji}(t)c_i)+\Pi_i(t) \ .
\end{equation}

The Duhamel integral gives for this system with initial condition
$c(t_0)$:

$$c(t)=U(t,t_0)c(t_0) + \int_{t_0}^t U(t,\tau)\Pi(\tau) \ \D
\tau \ , $$ where $\Pi(\tau)$ is the vector of fluxes with
components  $\Pi_i(\tau)$.

In particular, for stochastic in column operators $U(t,t_0)$
this formula gives: an identity for the linear conservation law
\begin{equation}\label{flux balance}
\sum_i c_i(t)= \sum_i c_i(t_0)+ \int_{t_0}^t \sum_i \Pi_i(\tau) \ \D \tau \ ,
\end{equation}
and an inequality for the $l_1$ norm
\begin{equation}\label{flux norm}
\|c(t)\| \leq \|U(t,t_0)c(t_0)\|+ \int_{t_0}^t \sum_i \| \Pi(\tau)\| \ \D \tau  \leq \|c(t_0)\|
 + \int_{t_0}^t \sum_i \| \Pi(\tau)\| \ \D \tau  \ .
\end{equation}
We need the last formula for the estimation of contraction
coefficients when the vector $c(t)$ is not positive.

\section{Kinetic Paths \label{KinPath}}

Two vertices are called adjacent if they share a common edge. A
directed path is a sequence of adjacent edges where each step
goes in direction of an edge. A vertex $A$ is {\it reachable}
from a vertex $B$, if there exists a directed path from $B$ to
$A$.

Formally, a {\em path} in a reaction graph is any finite
sequence of indexes (a multiindex) $I=\{i_1,i_2, \ldots i_q\}$
($q \geq 1$, $1\leq i_j\leq n$) such that $(i_k,i_{k+1}) \in
\mathcal{E}$ for all $k=1, \ldots , q-1$ (i.e. there exists a
reaction $A_{i_k} \to A_{i_{k+1}}$). The number of the vertices
$|I|$ in the path $I$ may be any natural number (including 1),
and any vertex $A_i$ can be included in the path $I$ several
times. If $q=1$ then we call the one-vertex path $I$
degenerated. There is a natural order on the set of paths:
$J>I$ if $J$ is continuation of $I$, i.e. $I=\{i_1,i_2, \ldots
i_q\}$ and $J=\{i_1,i_2, \ldots i_q, \ldots\}$. In this order,
the antecedent element (or the {\em parent}) for each $I$ is
$I^-$, the path which we produce from $I$ by deletion of the
last step. With this definition of parents $I^-$, the set of
the paths with a given start point is a rooted {\em tree}.

\begin{definition}\label{Definition 1} For each path $I=\{i_1,i_2, \ldots
i_q\}$ we define an auxiliary set of reaction, the {\em kinetic
path} $P_I$:
\begin{equation}\label{KinPathSch}
  \begin{CD}
B^I_{1(i_1)} @>{k_{i_2i_1}}>> B^2_{2(i_2)} @>{k_{i_3i_2}}>> \ldots @>{k_{i_qi_{q-1}}}>> B^I_{q(i_q)}\\
@VV{\kappa_{i_1\overline{i_2}}}V @VV{\kappa_{i_2\overline{i_3}}}V @. @VV{\kappa_{i_q}}V\\
  \end{CD}
\end{equation}
\end{definition}
The vertices $B^I_{l(i_l)}$ of the kinetic path
(\ref{KinPathSch}) are auxiliary components. Each of them is
determined by the path multiindex $I$ and the position in the
path $l$. There is a correspondence between the auxiliary
component $B^I_{l(i_l)}$ and the component $A_{i_l}$ of the
original network. The coefficient $\kappa_i$ is a sum of the
reaction rate coefficients for all outgoing reactions from the
vertex $A_i$ of the original network, and the coefficient
$\kappa_{i\overline{j}}$ is this sum without the term which
corresponds to the reaction $A_i \to A_j$:
 $$\kappa_i=\sum_{l,\ l\neq i} k_{li}, \;\; \kappa_{i\overline{j}}=\sum_{l,\ l \neq i,j}
 k_{li}\ .$$

A quantity, the concentration $b^I_{l(i_l)}$, corresponds to any
vertex of the kinetic path $B^I_{l(i_l)}$ and a kinetic equation of
the standard form can be written for this path. The end vertex, $
B^I_{q(i_q)}$, plays a special role in the further consideration and
we use the special notations: $i_I=i_q$, $A_I=A_{i_q}$,
$\varsigma_I= b^I_{q(i_q)}$,  $\kappa_I$ is the reaction rate
coefficient of the last outgoing reactions in (\ref{KinPathSch})
(the last vertical arrow) and $k_I$ is the reaction rate coefficient
of the last incoming reaction in (\ref{KinPathSch}) (the last
horizontal arrow).

We use $P_I^+$ for the incoming flux for the terminal vertex of the
kinetic path (\ref{KinPathSch}) and $P_I^-$ for the outgoing flux
for this vertex.

Let us consider the set $\mathcal{I}_1$ of all paths with the
same start point $i_1$ and the solutions of all the
correspondent kinetic equations with initial conditions:
$$b^I_{1(i_1)}=1, \; b^I_{l(i_l)}=0 \; {\rm for} \; l>1 \ .$$
For the concentrations of the terminal vertices this
self-consistent set of initial conditions gives the infinite
chain (or, to be more precise, the tree) of simple kinetic
equations for the set of variables $\varsigma_I$, $I\in
\mathcal{I}_1$:
\begin{equation}\label{ChainSimple}
\dot{\varsigma}_1=-\kappa_1(t) \varsigma_1, \;\dot{\varsigma}_I=-\kappa_I(t) \varsigma_I + k_I(t) \varsigma_{I^-} \ ,
\end{equation}
where index 1 corresponds to the degenerated path which consists of
one vertex (the start point only) and corresponds to $A_{i_1}$.

This simple chain of equations with initial conditions,
$\varsigma_1(t_0)=1$ and $\varsigma_I(t_0)=0$ for $|I|>1$, has
a recurrent representation of solution:
\begin{equation}\label{reccurentPathSolution}
\begin{split}
& \varsigma_1(t)=\exp\left(-\int_{t_0}^t \kappa_1(\tau) \, \D \tau\right), \;\; \\
& \varsigma_I(t)=\int_{t_0}^t  \exp\left(-\int_{\theta}^t \kappa_I(\tau) \, \D \tau\right)
 k_I (\theta) \varsigma_{I^-}(\theta) \, \D \theta \ .
\end{split}
\end{equation}

The analogues of the Kirchhoff rules from the theory of electric or
hydraulic circuits are useful for outgoing flux of a path
$J\in\mathcal{I}_1$ and for incoming fluxes of the paths which $I$
are the one-step continuations of this path (i.e. $I^-=J$):
\begin{equation}\label{Kirkhoff}
\kappa_J \varsigma_J=\sum_{I,\ I^-=J} k_I \varsigma_{I^-} \ .
\end{equation}
Let us rewrite this formula as a relation between the outgoing flux
$P_J^-$ from the last vertex of $J$ and incoming fluxes $P_I^+$ for
the last vertices of paths $I$ ($I^-=J$):
\begin{equation}\label{KirkhoffFlux}
P_J^-=\sum_{I,\ I^-=J} P_I^+ \ .
\end{equation}

The Kirchhoff rule (\ref{KirkhoffFlux}) together with the
kinetic equation for given initial conditions immediately
implies the following summation formula.

\begin{theorem}\label{Theorem 1} Let us consider the solution to the initial kinetic
equations (\ref{kinurT}) with the initial conditions
$c_{j}(t_0)=\delta_{j i_1}$. Then
\begin{equation}\label{summationConcentr}
c_j(t)=\sum_{I\in \mathcal{I}_1, \ i_I=j}\varsigma_I(t)
\end{equation}
\end{theorem}
{\em Proof.} To prove this formula let us prove that the sum
from the right hand side (i) exists (ii) satisfies the initial
kinetic equations (\ref{kinurT}) and (iii) satisfies the
selected initial conditions.

Convergence of the series with positive terms follows from the
boundedness of the set of the partial sums, which follows from the
Kirchhoff rules. According to them,
 $$\sum_{I\in \mathcal{I}_1} \varsigma_I(t)\equiv 1$$
because $\mathcal{I}_1$ consists of the paths with the selected
initial point $i_1$ only.

The sum $$C_j=\sum_{I\in \mathcal{I}_1, \ i_I=j}\varsigma_I$$
satisfies the kinetic equation (\ref{kinurT}). Indeed, let
$\mathcal{I}_{1j}=\{I\in \mathcal{I}_1 \ | \ i_I=j \}$ be the set of
all paths from $i_1$ to $j$. Let us find the set of all paths of the
form $\{I^- \ | \ I \in \mathcal{I}_{1j} \}$. This set (we call it
$\mathcal{I}_{1j}^-$) consists of all paths to all points which are
connected to $A_j$ by a reaction:
$$\mathcal{I}_{1j}^-= \bigcup_{(l,j)\in \mathcal{E}}
\mathcal{I}_{1l}\ .$$ From this identity and the chain of the kinetic equations
(\ref{ChainSimple}) we get immediately that
\begin{equation}\label{kinurTTT}
\frac{\D C_i}{\D t}=\sum_{j, \ j\neq i} (k_{ij}(t) C_j -
k_{ji}(t)C_i),
\end{equation}
The coincidence of the initial conditions for $c_i$ and $C_i$
is obvious. Hence, because of the uniqueness theorem for
equations (\ref{kinurT}) we proved that $c_i \equiv C_i$. $
\square$

It is convenient to reformulate Theorem~\ref{Theorem 1} in the
terms of the fundamental operator $U(t,t_0)$. The $i$th column
of $U(t,t_0)$ is a solution of (\ref{kinurT})
$c_j(t)=u_{ji}(t,t_0)$ $(j=1, \ldots ,n)$ with initial
conditions $c_j(t_0)=\delta_{ij}$. Therefore, we have proved
the following theorem. Let $\mathcal{I}_{ij}$ be the set of all
paths with the initial vertex $A_i$ and the end vertex $A_j$
and $\varsigma_I (t)$ be the solutions of the chain
(\ref{ChainSimple}) for $i_1=i$ with initial conditions:
$\varsigma_1(t_0)=1$ and $\varsigma_I(t_0)=0$ for $|I|>1$.

\begin{theorem}\label{Theorem 2}
\begin{equation}\label{summation formula}
u_{ji}(t,t_0)=\sum_{I\in \mathcal{I}_{ij}} \varsigma_I(t) \ . \ \ \ \ \ \square
\end{equation}
\end{theorem}

{\bf Remark 1.} It is important that all the terms in the sum
(\ref{summation formula}) are non-negative, and any partial sum
gives the approximation to $u_{ji}(t,t_0)$ from below.

{\bf Remark 2.} If the kinetic coefficients are constant then
the Laplace transform gives a very simple representation for
solution to the chain (\ref{ChainSimple}) (see also
computations in  \cite{Flomenbom2005,Sun2006}). The kinetic
path $I$ (\ref{KinPathSch}) is a sequence of elementary links
\begin{equation}\label{link}
  \begin{CD}
\ldots @>{k_{i_ri_{r-1}}}>> B^r_{r(i_r)} @>{k_{i_{r+1}i_r}}>> \ldots \\
@.                  @VV{\kappa_{i_r\overline{i_{r+1}}}}V @. \\
  \end{CD}
\end{equation}
The transfer function $W_{i_r}(p)$ for the link (\ref{link}) is
the ratio of the output Laplace Transform to the input Laplace
Transform for the equation. Let the input be a function
$X_{i_r}(t)$ and the output be $Y_{i_r}(t)=b_{i_r}(t)$, where
$b_{i_r}(t)$ is the solution to equation
$$\dot{b}_{i_1}=-\kappa_{i_1}{b}_{i_r}+X_{i_1}(t)\, ; \;
\dot{b}_{i_r}=-\kappa_{i_r}{b}_{i_r}+k_{i_ri_{r-1}}X_{i_r}(t) \; (r>1)$$
with zero initial conditions. The Laplace transform gives
$$W_{i_1}=\frac{1}{p+\kappa_{i_1}}\, , \;\;
W_{i_r}=\frac{k_{i_{r}i_{r-1}}}{p+\kappa_{i_r}}\; (r>1)$$ for a link (\ref{link}) and for
the whole path (\ref{KinPathSch}) we get
\begin{equation}\label{TransferPath}
W_{I}=\frac{1}{p+\kappa_{i_1}}\prod_{r=2}^q \frac{k_{i_{r}i_{r-1}}}{p+\kappa_{i_r}} \, .
\end{equation}
(compare, for example, to formula (9) in \cite{Sun2006}). It is
worth to mention commutativity of this product: it does not
change after a permutation of internal links. For the infinite
chain (\ref{ChainSimple}) with initial conditions,
$\varsigma_1(0)=1$ and $\varsigma_I(0)=0$ for $|I|>1$, the
Laplace transformation of solutions is
\begin{equation}
\mathcal{L} \varsigma_I =W_{I}
\end{equation}

\section{Evaluation of Ergodicity Coefficient \label{sec3}}

\subsection{Preliminaries: Weak Ergodicity and Annihilation
Formula}

\subsubsection{Geometric Criterion of Weak Ergodicity}

In this Subsection, let us consider a reaction kinetic system
(\ref{kinurT}) with constant coefficients $k_{ji}>0$ for $(i,j)
\in \mathcal{E}$.

A set $E$ is {\it positively invariant} with respect to the
kinetic equations (\ref{kinurT}), if any solution $c(t)$ that
starts in $E$ at time $t_0$ ($c(t_0) \in E$) belongs to $E$ for
$t>t_0$ ($c(t) \in E$ if $t>t_0$). It is straightforward to
check that the standard simplex $\Sigma = \{c\,| \, c_i \geq 0,
\, \sum_i c_i =1\}$ is a positively invariant set for kinetic
equation (\ref{kinurT}): just check that if  $c_i=0$ for some
$i$, and all $c_j \geq 0$ then $\dot{c}_i \geq 0$. This simple
fact immediately implies the following properties of ${K}$:
\begin{itemize}
\item{All eigenvalues $\lambda$ of ${K}$ have non-positive
    real parts, $Re \lambda \leq 0$, because solutions
    cannot leave $\Sigma$ in positive time;}
\item{If $Re \lambda = 0$ then $\lambda = 0$, because
    the intersection of $\Sigma$ with any plane is a polygon,
    and a polygon cannot be invariant with respect to
    rotations to sufficiently small angles;}
\item{The Jordan cell of ${K}$ that corresponds to the zero
    eigenvalue is diagonal -- because all solutions should
    be bounded in $\Sigma$ for positive time.} \item{The
    shift in time operator $\exp({K} t)$ is a contraction
    in the $l_1$ norm for $t>0$: there exists such a
    monotonically decreasing (non-increasing) function
    $\delta(t)$ ($t>0$, $0<\delta(t)\leq 1$,  that for any
    two solutions of (\ref{kinurT}) $c(t), c'(t) \in
    \Sigma$
\begin{equation}\label{contractiondefinition}
\sum_i |c_i(t) -c'_i(t)| \leq \delta(t) \sum_i |c_i(0) -c'_i(0)|.
\end{equation} }
\end{itemize}
Moreover, if for $c(t), c'(t) \in \Sigma$  the values of all
linear conservation laws coincide then $\sum_i |c_i(t)
-c'_i(t)| \to 0$ monotonically when $t \to \infty$.

The first-order kinetic system is {\em weakly ergodic} if it allows
only the conservation law: the sum of concentration. Such a system
forgets its initial condition: distance between any two trajectories
with the same value of the conservation law tends to zero when time
goes to infinity.

The difference between weakly ergodic and ergodic systems is in
obligatory existence of a strictly positive stationary distribution:
for an ergodic system, in addition, a strictly positive steady state
exists: $Kc=0$ and all $c_i>0$. Examples of weakly ergodic but not
ergodic systems: a chain of reactions $A_1 \to A_2 \to \ldots \to
A_n$ and symmetric random walk on an infinite lattice.

The weak ergodicity of the network follows from its topological
properties.

\begin{theorem}\label{Theorem 3}  The following properties are equivalent (and each one of
them can be used as an alternative definition of weak
ergodicity):
\begin{enumerate}
\item{There exists a unique independent linear conservation
    law for kinetic equations (this is $b^0(c)=\sum_i c_i =
    {\rm const}$).}
 \item{For any normalized initial state $c(0)$ ($b^0(c)=1$)
     there exists a limit state $$c^*= \lim_{t\rightarrow
     \infty } \exp(Kt) \, c(0)$$ that is the same for all
     normalized initial conditions: For all $c$,
     $$\lim_{t\rightarrow \infty } \exp(Kt) \, c = b^0(c)
     c^*.$$}
\item{For each two vertices  $A_i, \: A_j \: (i \neq j)$ we
    can find such a vertex $A_k$ that is reachable both
    from $A_i$ and from $A_j$. This means that the
    following structure exists:
\begin{equation}\label{elementBridge}
    A_i \to \ldots \to A_k \leftarrow \ldots \leftarrow
A_j \ .
\end{equation}
One of the paths can be degenerated: it may be $i=k$ or
$j=k$.}
\item{For $t>0$ operator $\exp(Kt)$ is a strong contraction
    in the invariant subspace $\sum_i c_i=0$ in the $l_1$
    norm: $\|\exp(Kt) x\| \leq \delta(t) < 1$, the function
    $\delta(t)>0$ is strictly monotonic and $\delta(t) \to
    0$ when $t \to \infty$\;\;\; $\square$.}
\end{enumerate}
\end{theorem}

The proof of this theorem could be extracted from detailed
books about Markov chains and networks
(\cite{MeynNets2007,VanMieghem2006}). In its present form it
was published in \cite{Ocherki} with explicit estimations of
the ergodicity coefficients.

Let us demonstrate how to prove the geometric criterion of weak
ergodicity, the equivalence $1 \Leftrightarrow 3$.

Let us assume that there are several linearly independent
conservation laws, linear functionals $b^0(c),b^1(c), \ldots ,
b^m(c)$, $m \geq 1$. The linear transform $c \mapsto (b^1(c),\ldots
,b^m(c))$ maps the standard simplex $\Sigma_n$ in $\mathbb{R}^n$
($c_i \geq 0$, $\sum_i c_i=1$) onto a polyhedron $D \subset
\mathbb{R}^m$. Because of linear independence of the system
$b^0(c),b^1(c), \ldots , b^m(c)$, $m \geq 1$, this $D$ has nonempty
interior. Hence, it has no less than $m+1$ vertices $w_1,\ldots,
w_q$, $q >m$.

The preimage of every point $x\in D$ in $\Sigma_n$ is a positively
invariant subset with respect to kinetic equations because the
standard simplex is positively invariant and the functionals
$b^i(c)$ are the conservation laws. In particular, preimage of every
vertex $w_q$ is a positively invariant face of $\Sigma_n$, $F_q$;
$F_q \cap F_r = \emptyset$ if $q\neq r$.

Each vertex $v_i$ of the standard simplex corresponds to a component
$A_i$: at this vertex $c_i=1$ and other $c_j=0$ there. Let the
vertices from $F_q$ correspond to the components which form a set
$S_q$; $S_q \cap S_r =\emptyset$ if $q\neq r$.

For any $A_i \in S_q$ and every reaction $A_i \to A_j$ the component
$A_j$ also belongs to $S_q$ because $F_q$ is positively invariant
and a solution to kinetic equations cannot leave this face.
Therefore, if $q\neq r$, $A_i \in S_q$ and $A_j \in S_r$ then there
is no such vertex $A_k$ that is reachable both from $A_i$ and from
$A_j$. We proved  the implication $3\Rightarrow 1$.

Now, let us assume that the statement 3 is wrong and there
exist two such components $A_i$ and $A_j$ that no components
are reachable both from $A_i$ and $A_j$. Let $S_i$ and $S_j$ be
the sets of components reachable from $A_i$ and $A_j$
(including themselves), respectively; $S_i \cap S_j
=\emptyset$.

For every concentration vector $c \in \mathbb{R}^n$ a limit exists
$c^*(c)=\lim_{t\to \infty}\exp(Kt)\ c$ (because all eigenvalues of
$K$ have non-positive real part and the Jordan cell of ${K}$ that
corresponds to the zero eigenvalue is diagonal). The operator $c
\mapsto c^*(c) $ is linear operator in $\mathbb{R}^n$. Let us define
two linear conservation laws:
$$b^i(c)=\sum_{A_r\in S_i}c_r^*(c), \ \ b^j(c)=\sum_{A_r\in
S_j}c_r^*(c) \ .$$ These functionals are linearly independent
because for a vector $c$ with coordinates $c_r=\delta_{ri}$ we
get $b^i(c)=1$, $b^j(c)=0$ and for a vector $c$ with
coordinates $c_r=\delta_{rj}$ we get $b^i(c)=0$, $b^j(c)=1$.
Hence, the system has at least two linearly independent linear
conservation laws.  Therefore, $1\Rightarrow 3$.

\subsubsection{Annihilation Formula}

Let us return to general time--dependent kinetic equations
(\ref{kinurT}).

In this Section, we find an exact expression for  the contraction
coefficients $\delta(t,t_0)$ for the time evolution operator
$U(t,t_0)$ in $l_1$ norm on the invariant subspace $\{x\ | \ \sum_i
x_i=0\}$. The unit $l_1$-ball in this subspace is a polyhedron with
vertices $g^{ij}=\frac{1}{2}(e^i-e^j)$, where $e_i$ are the standard
basic vectors in $\mathbb{R}^n$ (\ref{ballVert}). The contraction
coefficient of an operator $U$ is its norm on that subspace
(\ref{normSubSp}), this is half of the maximum of the $l_1$
distances between columns  of $U$.

The kinetic path summation formula (\ref{summation formula})
estimates the matrix elements of $U(t,t_0)$ from below, but this
does not give the possibility to evaluate the difference between
these elements. To use the summation formula efficiently, we need
another expression for the contraction coefficient.

The $i$th column of $U(t,t_0)$ is a solution of the kinetic
equations (\ref{kinurT}) $c_j(t)=u_{ji}(t,t_0)$ $(j=1, \ldots ,n)$
with initial conditions $c_j(t_0)=\delta_{ij}$. For each $j$ let us
introduce the incoming flux for the vertex $A_j$ in this solution:
$$\Pi_j^i(t)=\sum_q k_{jq}(t) c_q(t)$$ (the upper index indicates
the number of column in $U(t,t_0)$, the lower index corresponds to
the number of vertex $A_j$).

Formula (\ref{normSubSp}) for the contraction coefficient gives

$$\delta(t,t_0)=\frac{1}{2}\max_{i,j}\|U(t,t_0)(e^i-e^j)\| \ .$$

$U(t,t_0)(e^i-e^j)$ is a solution to the kinetic equation
(\ref{kinurT}) with initial conditions $c_i(t_0)=1$, $c_j(t_0)=-1$
and $c_q(t_0)=0$ for $q\neq i,j$. This is the difference between two
solutions, $c^+_q(t)=u_{qi}(t,t_0)$ and $c^-_q(t)=u_{qj}(t,t_0)$.
Let us use the notation

$$G^{ij}(t)=\frac{1}{2}U(t,t_0)(e^i-e^j) \ .$$

For each $q$ we define
 $$\Pi^+_q= \sum_{l, c^+_l>c^-_l} k_{ql} (c^+_l-c^-_l), \;\;
   \Pi^-_q= \sum_{l, c^+_l<c^-_l} k_{ql} (c^-_l-c^+_l), \;\; \Pi^{\pm}_q \geq 0 \ .$$
The decrease in the $l_1$ norm of $c^+(t)-c^-(t)$ can be represented
as an annihilation of a flux $\Pi^{\pm}_q(t)$ with an equal amount
of concentration $c^+(t)-c^-(t)$ from the vertex $A_q$ by the
following rules:
\begin{enumerate}
\item{If $c_q=c^+_q(t)-c^-_q(t)>0$ then the flux $\Pi^-_q$
    annihilates with an equal amount of positive concentration
    stored at vertex $A_q$ (Fig.~\ref{NegAnnih});}
\item{If $c_q=c^+_q(t)-c^-_q(t)<0$ then the flux $\Pi^+_q$
    annihilates with an equal amount of negative concentration
    stored at vertex $A_q$ (Fig.~\ref{PosAnnih});}
\item{If $c_q=c^+_q(t)-c^-_q(t)=0$ then the flux
    $\min\{\Pi^+_q, \Pi^-_q\}$ annihilates with the equal
    amount from the opposite flux (Fig.~\ref{MinAnnih}).}
\end{enumerate}

Let us summarize these rules in one formula:

\begin{figure}
\centering{
\subfloat[$c>0$, the negative flux annihilates]{\label{NegAnnih}\includegraphics[width=0.25\textwidth]{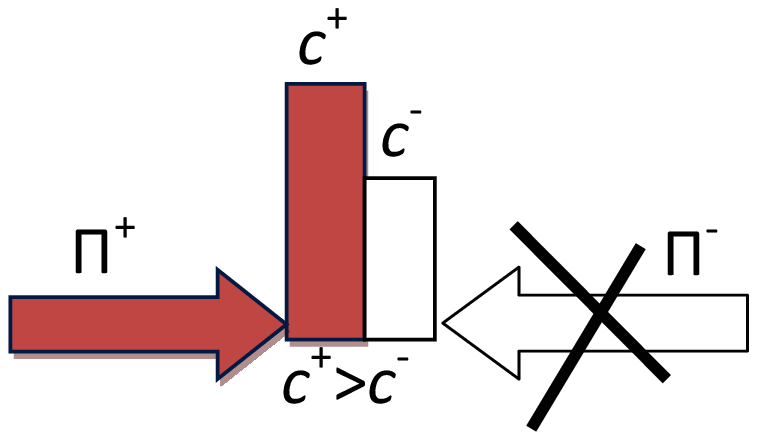}} \qquad
\subfloat[$c<0$, the positive flux
annihilates]{\label{PosAnnih}\includegraphics[width=0.25\textwidth]{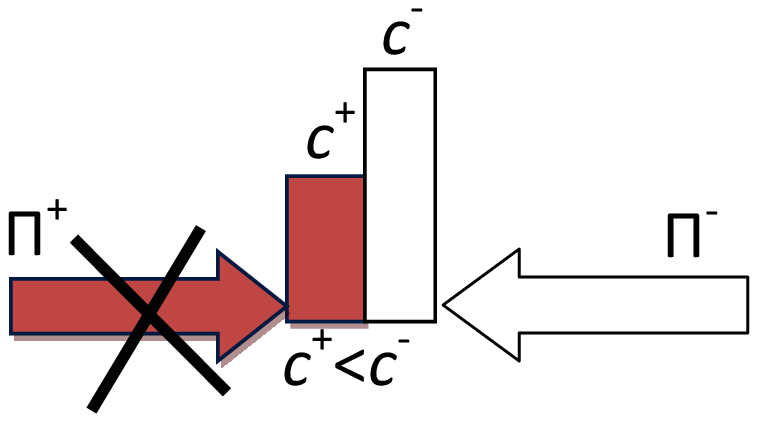}} \qquad
\subfloat[$c=0$, the minimal flux
annihilates]{\label{MinAnnih}\includegraphics[width=0.25\textwidth]{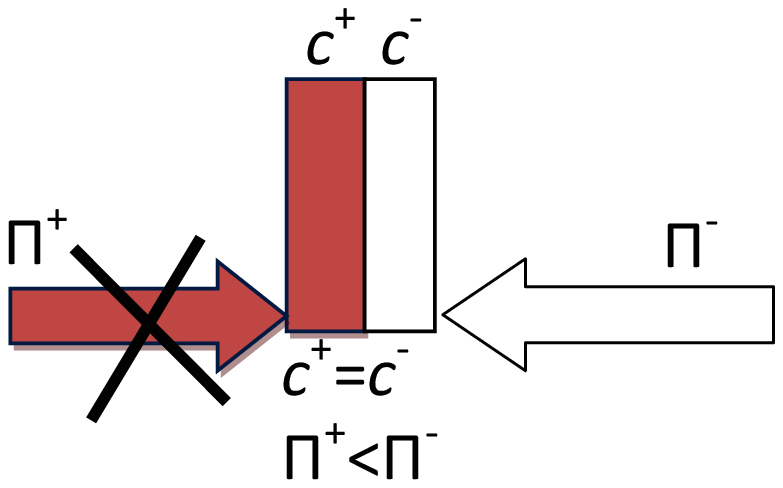}}
}\caption{Annihilation of fluxes. \label{fluxes}}
\end{figure}

\begin{proposition}\label{Proposition 1}
\begin{equation}\label{l1Deriv}
\begin{split}
\frac{\D}{\D t} \|G^{ij}(t) \|_{l_1} =& -\sum_{q,\ c^+_q>c^-_q }
\Pi^-_q(t) - \sum_{q,\ c^+_q<c^-_q } \Pi^+_q(t) \\ &- \sum_{q,\
c^+_q=c^-_q } \min\{\Pi^+_q(t),\Pi^-_q(t)\} \ . \;\;\; \square
\end{split}
\end{equation}
\end{proposition}

Immediately from (\ref{l1Deriv}) we obtain the following
integral formula
\begin{equation}\label{l1Integr}
\begin{split}
1-\|G^{ij}(t) \|_{l_1} = &\int_{t_0}^t \left(\sum_{q,\ c^+_q>c^-_q }
\Pi^-_q(\tau) + \sum_{q,\ c^+_q<c^-_q } \Pi^+_q(\tau) \right. \\&+
\left. \sum_{q,\ c^+_q=c^-_q }
\min\{\Pi^+_q(\tau),\Pi^-_q(\tau)\})\right)\ \D \tau \ .
\end{split}
\end{equation}

The annihilation formula gives us a better understanding of the
nature of contraction but is not fully constructive because it uses
fluxes from solutions to the initial kinetic equation
(\ref{kinurT}).

\subsection{Multi--Sheeted Extensions of Kinetic System \label{Sec:MultiSheeted}}

Let us introduce a multi--sheeted extension of a kinetic system.

\begin{definition}\label{Definition 2} The vertices of a {\em multi--sheeted
extension} of the system \eqref{kinurT} are $\mathcal{A}\times
K$ where $K$ is a finite or countable set. An individual vertex
is $(A_i,l)$ ($l\in K$). The corresponding concentration is
$c_{(i,l)}$. The reaction rate constant for $(A_i,l) \to
(A_j,r)$ is $k_{(j,r)(i,l)}\geq 0$. This system is a
multi--sheeted extension of the initial system if an identity
holds:
\begin{equation}\label{manySheet}
\sum_r k_{(j,r)(i,l)}=k_{ji} \ \mbox{ for all } \ l \ .
\end{equation}
\end{definition}
This means that the flux from each vertex is distributed
between sheets, but the sum through sheets is the same as for
the initial system. We call the kinetic behavior of the sum
$c_i=\sum_lc_{(i,l)}$ the {\em base kinetics}.

A simple proposition is important for further consideration.

\begin{figure}
\centering{
\includegraphics[width=0.8\textwidth]{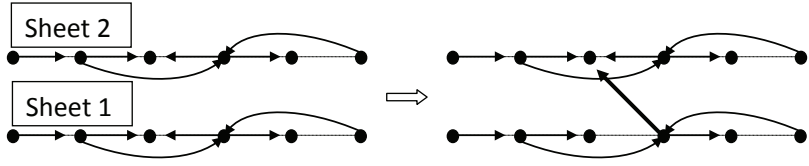}
}\caption{Redirection of a reaction from one sheet to another with preservation of the base kinetics.
The redirected reaction is highlighted by bold. \label{redir}}
\end{figure}

\begin{proposition}\label{Proposition 2} If $c_{(i,l)}(t)$ is a solution to
the extended multi--sheeted system then
\begin{equation}\label{MSheetSum}
c_i(t)=\sum_l c_{(i,l)}(t)
\end{equation}
is a solution to the initial system and
\begin{equation}\label{MSheetNorm}
\sum_{il}|c_{(i,l)}(t)| \geq \sum_i |c_i(t)| \ .
\end{equation}
(Here we do not assume positivity of all $c_i$). \ \ \ \
$\square$
\end{proposition}

Formula (\ref{manySheet}) allows us to redirect reactions from
one sheet to another (Fig.~\ref{redir}) without any change of
the base kinetics. In the next section we show how to use this
possibility for effective calculations.

Formula (\ref{MSheetSum}) means that kinetics of the extended
system in projection on the initial space is the base kinetics:
the components $(A_i,l)$ are projected in $A_i$ the projected
vector of concentrations is $c_i=\sum_lc_{(i,l)}$ and the
projected kinetics is given by the initial Master equation with
the projected coefficients $k_{ji}=\sum_r k_{(j,r)(i,l)}$.
``Recharging" is any change of the non-negative extended
coefficients $k_{(j,r)(i,l)}$ which does not change the
projected coefficients.

The key role in the further estimates plays formula
(\ref{MSheetNorm}). We will apply this formula to the solutions
with the zero sums of coordinates, they are differences between
the normalized positive solutions.

\subsection{Fluxes and Mixers}

In this Subsection, we present the system of estimates for the
contraction coefficient. The main idea is based on the
following property which can be used as an alternative
definition of weak ergodicity (Theorem~\ref{Theorem 3}): For
each two vertices $A_i, \: A_j \: (i \neq j)$ we
    can find a vertex $A_q$ that is reachable both
    from $A_i$ and from $A_j$. This means that the
    following structure exists:
 $$A_i \to \ldots \to A_q \leftarrow \ldots \leftarrow A_j.$$
One of the paths can be degenerated: it may be $i=q$ or $j=q$.
The positive flux from $A_i$ meets the negative
flux from $A_j$ at point  $A_q$ and one of them annihilates with the equal
amount of the concentration of opposite sign.

Let us generalize this construction. Let us fix three different
vertices: $A_i$ (the ``positive source"), $A_j$ (the ``negative
source") and $A_q$ (the ``mixing point"). The degenerated case
$q=i$ or $q=j$ we discuss separately. Let $S^+$ be such a
system of vertices that $A_i\in S^+$, $A_q \notin S^+$  and
there exists an oriented path in $S^+\cup \{A_q\}$ from $A_i$
to $A_q$. Analogously, let $S^-$ be such a system of vertices
that $A_j\in S^-$, $A_q \notin S^-$  and there exists an
oriented path in $S^-\cup \{A_q\}$ from $A_j$ to $A_q$. We
assume that $S^+\cap S^- =\emptyset$.

With each subset of vertices $S$ we associate a kinetic system
(subsystem): for $A_r \in S$
\begin{equation}\label{subsystem}
\dot{c}_r=\sum_{l, \ A_l \in S, \ r\neq l} k_{rl} c_l - \sum_{p=1}^n
k_{pr}c_r \ .
\end{equation}
In this subsystem, we retain all the outgoing reaction for $A_r \in
S$ and delete the reactions which lead to vertices in $S$ from
``abroad".

The flux $\Pi_S^+$ from $S^+$ to $A_q$ is $$\Pi_S^+=\sum_{r,\ A_r
\in S^+} k_{qr} c_r(t) \ ,$$ where $c_r(t)$ is a component of the
solution of (\ref{subsystem}) for $S=S^+$ with initial conditions
$c_r(t_0)=\delta_{ri}$. Analogously, we define the flux
$$\Pi_S^-=\sum_{r,\ A_r \in S^-} k_{qr} c_r(t) \ , $$ where $c_r(t)$
is a component of the solution of (\ref{subsystem}) for $S=S^-$ with
initial conditions $c_r(t_0)=\delta_{rj}$.  Decrease of the norm
$\|G^{ij}(t)\| $ is estimated by the following theorem.

The system $S^+, S^-, A_q$ we call a {\em mixer}, that is a
device for mixing. An {\em elementary  mixer}  consists of two
kinetic paths $A_i \to \ldots \to A_q \leftarrow \ldots
\leftarrow A_j$ (\ref{elementBridge}) with the corespondent
outgoing reactions:

\begin{equation}\label{ElementaryMixer}
  \begin{CD}
A_{i_1} @>{k_{i_2i_1}}>> \ldots @>{k_{i_ri_{r-1}}}>> A_{i_r} @<{k_{i_r i_{r+1}}}<< \ldots @<{k_{i_{r+l-1}i_{r+l}}}<< A_{i_{r+l}}\\
@VV{\kappa_{i_1\overline{i_2}}}V @. @VV{\kappa_{i_r}}V@. @V{\kappa_{i_{r+l}{\overline{i_{r+l-1}}}}}VV\\
  \end{CD}
\end{equation}
where $i_1=i$, $i_r=q$, $i_{r+l}=j$.

The degenerated elementary mixer consists of one kinetic path:
\begin{equation}\label{ElementaryMixerDegen}
  \begin{CD}
A_{i_1} @>{k_{i_2i_1}}>> A_{i_2} @>{k_{i_3i_2}}>> \ldots @>{k_{i_ri_{r-1}}}>> A_{i_r}\\
@VV{\kappa_{i_1\overline{i_2}}}V @VV{\kappa_{i_2\overline{i_3}}}V @. @VV{\kappa_{i_r}}V\\
  \end{CD}
\end{equation}
where $i_1=i$, $i_r=j$.

\begin{theorem}\label{Theorem 4}
\begin{equation}\label{subsystemsEstim}
\|G^{ij}(t)\| \leq 1- \int_{t_0}^t \min\{\Pi_S^+,\Pi_S^-\} \ \D t \ .
\end{equation}
\end{theorem}

\begin{figure}
\centering{
\includegraphics[width=0.8\textwidth]{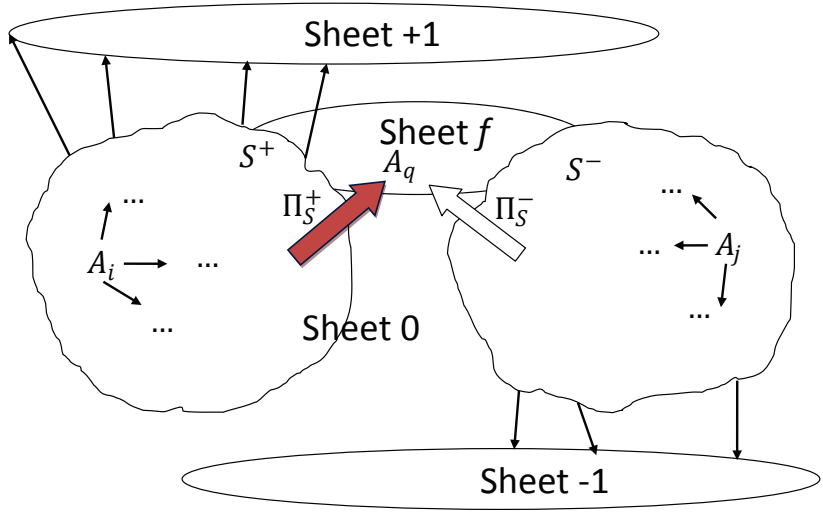}
}\caption{A mixer: two subsystems, $S^+$ (includes $A_i$) and $S^-$ (includes $A_j$).
There may be outgoing reactions from $S^{\pm}$ but all incoming reactions to $S^{\pm}$
from outside are deleted. A mixing point $A_q$ and two fluxes,
positive from $S^+$ (marked by dark color) and negative from $S^-$, meet at the mixing point. \label{Subsys}}
\end{figure}

{\em Proof.} To prove this theorem let us organize a 4--sheeted
extension of the initial kinetic system as it is demonstrated
in Fig.~\ref{Subsys}. Subsystems $S^{\pm}$ including the
positive source (initial concentration $+1$) and the negative
source (initial concentration $-1$) belong to level 0.
Reactions from $S^{\pm}$ to $A_q$ are redirected to the sheet
$f$, reactions from $S^+$ to other vertices, which do not
belong to $S^+$, go to sheet $+1$, reactions from $S^-$ to
other vertices, which do not belong to $S^-$, go to sheet $-1$.
The incoming flux to the sheet $f$ is $\Pi_S^+-\Pi_S^-$.

Let us introduce the following notations:

$$C_S^+=\sum_{A_p \in S^+}c_{(p,0)}+\sum_{q=1}^n c_{(q,1)} \ ;$$
$$C_S^-=-\sum_{A_p \in S^-}c_{(p,0)}-\sum_{q=1}^n c_{(q,-1)} \ ;$$
$$C_f=\sum_{r=1}^n |c_{(r,f)}| \ . $$

We consider solution to the kinetic equations for the multi--sheeted
system with initial conditions: $c_{(i,0)}(t_0)=1$,
$c_{(j,0)}(t_0)=-1$ and all other concentrations are equal to zero
at time $t_0$. In this case, some of the signs of concentrations are
known for $t \geq t_0$ due to the organization of the redirection of
reactions (Fig.~\ref{Subsys}):
\begin{equation}
\begin{split}
&c_{(p,0)}\geq 0 \ \ \mbox{for} \ \ A_p \in S^+ \ ,\ \ c_{(p,0)}\leq 0 \ \ \mbox{for} \ \ A_p \in S^- \ , \\ & c_{(p,0)}= 0 \ \ \mbox{for} \ \ A_p \notin S^+\cup S^- \ ,\\
& c_{(q,1)} \geq 0, \ \ c_{(q,-1)} \leq 0 \ .
\end{split}
\end{equation}

Let us use (\ref{flux balance}) for $S^+$ with the sheet $+1$ and for $S^-$ with the sheet $-1$. We get immediately
\begin{equation}\label{ConcDerSheetspm1}
\frac{\D C_S^+}{\D t}=\Pi_S^+ \ , \ \ \frac{\D C_S^-}{\D t}=\Pi_S^-
\end{equation}
Analogously, we can use (\ref{flux norm}) for the sheet $f$ and get
\begin{equation}\label{ConcDerSheetf}
\frac{\D C_f}{\D t}\leq |\Pi_S^+-\Pi_S^-| \ .
\end{equation}
For the norm of the base vector of concentration $c$ the
inequality holds (Proposition~\ref{Proposition 2}):
$$\|c\| \leq C^+_S+C^-_S +C_f \ .$$

Finally, we combine this inequality with (\ref{ConcDerSheetspm1}), (\ref{ConcDerSheetf}) and get
$$\|c(t)\| \leq 2 -2 \int_{t_0}^t \min\{\Pi_S^+(\tau),\Pi_S^-(\tau)\} \ \D \tau \ \ \ \ \square$$

For the degenerate case the path from $A_i$ goes directly to
$A_j$ (or inverse). let us assume that there is a subsystem
$S^+$, $A_i \in S^+$, the mixing point $A_q$ coincides with
$A_j$ and the flux $\Pi^+_S$ is
$$ \Pi_S^+=\sum_{r,\ A_r \in S^+} k_{jr} c_r(t) \ ,$$ where $c_r(t)$ is a
component of the solution of (\ref{subsystem}) for $S=S^+$ with initial
conditions $c_r(t_0)=\delta_{ri}$.

\begin{theorem}\label{Theorem 5}
\begin{equation}\label{subsystemsEstimDeg}
\|G^{ij}(t)\| \leq 1- \int_{t_0}^{\min\{ t, t_1\}} {\Pi_S^+(\tau)} \ \D \tau ,
\end{equation}
where $\kappa_j=\sum_p k_{pj}$ and $t_1$ is a solution to
equation
\begin{equation}\label{stopflux}
\int_{t_0}^t {\Pi_S^+}(\tau) \exp(-\kappa_j (t-\tau)) \ \D \tau = \exp(-\kappa_j t) \
.
\end{equation}
\end{theorem}

{\em Proof.} This theorem is also proved by the construction of
the appropriate multi--sheeted extension of the kinetic system.
For the degenerated case we need only two additional sheets:
subsystem $S^+$ including the positive source $A_i$ (initial
concentration $+1$) and the negative source $A_j$ (initial
concentration $-1$) belong to level 0. Reactions from $S^+$ to
other vertices, which do not coincide with $A_j$, go to sheet
$+1$, reactions from $A_j$  to other vertices go to sheet $-1$.
The concentration of $A_{(j,0)}$ is
$$c_{(j,0)}(t)=\int_{t_0}^t {\Pi_S^+}(\tau) \exp(-\kappa_j (t-\tau))
\ \D \tau - \exp(-\kappa_j t) \ .$$

Let us introduce the following notation:
$$C_S^+=\sum_{A_p \in S^+}c_{(p,0)}+\sum_{q=1}^n c_{(q,1)} \ ;$$
$$C^-=-c_{(j,0)}-\sum_{q=1}^n c_{(q,-1)} \ .$$

For $t \leq t_1$ concentrations $c_{(j,0)}(t)$ and all $c_{(q,-1)}$ are negative, hence \begin{equation}
\frac{\D C_S^+}{\D t}=\frac{\D C^-}{\D t}=-\Pi_S^+(t)
\end{equation}
and for the norm of the correspondent solution for the base system we get the inequality
\begin{equation}
\|c(t)\|\leq 2-2\int_{t_0}^{\min\{ t, t_1\}} {\Pi_S^+(\tau)} \ \D \tau \ \ \ \ \ \square
\end{equation}

The kinetic path summation formula gives us a family of
estimates of $\Pi_S^{\pm}$ from below. For each pair $i,j$ we
can find the best of available estimates of $\|G^{ij}(t)\|$
(the smallest one for various choices of $A_q$ and subsets
$S^{\pm}$) and then among all pairs of $i,j$ we should choose
the ``most pessimistic" evaluation of $\|G^{ij}(t)\|$ (the
biggest one). It will give the evaluation of the contraction
coefficient from above.

\section{Simple example: Irreversible Cycle}

Let us demonstrate all results for a simple kinetic system, a
simple irreversible cycle:
\begin{equation}\label{cycle}
A_1 \xrightarrow{k_1} A_2 \xrightarrow{k_2} \ldots \xrightarrow{k_{n-1}} A_n \xrightarrow{k_{n}} A_1
\end{equation}
All $k_i>0$ and are constant in time. For enumeration of $A_i$
we use the standard cyclic order (mod$n$): $A_{n+j}\equiv A_j$.

The kinetic equations for this system are: $\dot{c} = Kc$ or
\begin{equation}
\frac{\D }{\D t}\left[\begin{array}{l}
c_1 \\
c_2 \\
\vdots\\
c_n
\end{array}\right]=
\left[\begin{array}{llll}
-k_1 &  0  & \ldots & k_n \\
 k_1 & -k_2& \ldots & 0 \\
\vdots &\vdots&\vdots&\vdots \\
0  &  0  &  0  & -k_n
\end{array}\right]\,
\left[\begin{array}{l}
c_1 \\
c_2 \\
\vdots\\
c_n
\end{array}\right]
\end{equation}
The characteristic equation for this system is $$\prod_{i=1}^n
(k_i+\lambda)=\prod_{i=1}^n k_i\, .$$ One eigenvalue for matrix
$K$ is, obviously, $\lambda=0$, the correspondent left
eigenvector is the linear conservation law $l_1=(1,1,\ldots
,1)$. The right eigenvector for this $\lambda$ is the steady
state
$r_1=\frac{1}{\sum_i\frac{1}{k_i}}(\frac{1}{k_1},\frac{1}{k_2},
\ldots , \frac{1}{k_n})^{\rm T}$ (normalized for $l_1 r_1=1$).
Other $n-1$ roots of the characteristic equations have strictly
negative real parts, $Re \lambda_i <0$ ($i>1$) and, in general,
cannot be found explicitly. For a given eigenvalue $\lambda$,
the eigenvectors have a simple structure:
\begin{equation}\label{eigenvectorscycle}
l_{{\lambda}\,i+1}=l_{{\lambda}\,i}\frac{\lambda+k_i}{k_i}\,
\;\;r_{{\lambda}\,i}=\frac{\psi_{{\lambda}\,i}}{k_i}\, , \,\,
\psi_{{\lambda}\,i-1}=\psi_{{\lambda}\,i}\frac{\lambda+k_i}{k_i}\,
.
\end{equation}

With the normalization condition: for eigenvalues $\lambda$,
$\lambda'$: $l_{\lambda}r_{\lambda'}=\delta_{\lambda
\lambda'}$, that is 1 for $\lambda = \lambda'$ and 0 for
$\lambda \neq \lambda'$.

 Two limit cases allow explicit analysis
of eigenvalues and eigenvectors of $K$:
\begin{enumerate}
\item{Systems with limiting steps: one constant is much
    smaller than others, let it be $k_n$, $k_n\ll k_i$,
    ($i=1,\ldots , n-1$);}
\item{Fully symmetric systems, $k_1=k_2=\ldots=k_n$.}
\end{enumerate}
For systems with limiting steps ($k_n\ll k_i$, ($i=1,\ldots ,
n-1$)) the eigenvalues are close to $-k_1, \ldots , - k_{n-1}$
and the relaxation is limited by the second constant, the next
to the minimal one (detailed analysis is provided in
\cite{GorbaRadul2008,GorbanRadZinChemEngSci2010}).

For a symmetric system ($k_1=k_2=\ldots=k_n=k$),  the
eigenvalues are: $\lambda_q=k \exp\left(\frac{2\pi i
q}{n}\right)-1$ for $q=1,\ldots, n$. There are $n$ distinct
eigenvalues, one of them, $\lambda_n=0$, the other have
negative real part: $Re \lambda_q = k\left[\cos\left(\frac{2\pi
i q}{n}\right)-1\right]$. Let us further take $k=1$ for this
system (include $k$ into dimensionless time). For the left and
right eigenvectors (\ref{eigenvectorscycle}) we have two waves
moving in opposite directions, $l_{q\,j+1}=l_{qj}
\exp\left(\frac{2\pi i q}{n}\right)$, $r_{q\,j-1}=r_{q\,j}
\exp\left(\frac{2\pi i q}{n}\right)$. We can take with respect
to the normalization condition, $l_q r_p=\delta_{qp}$:
\begin{equation}\label{eigenvectorscycleSYM}
\begin{split}
& l_{q}=\left(1,\exp\left(\frac{2\pi i q}{n}\right),
\exp\left(2\frac{2\pi i q}{n}\right),
\ldots , \exp\left((n-1)\frac{2\pi i q}{n}\right) \right)\, ,\\
&r_{q}=\frac{1}{n}\left(1,\exp\left(-\frac{2\pi i q}{n}\right),
\exp\left(-2\frac{2\pi i q}{n}\right),
\ldots , \exp\left(-(n-1)\frac{2\pi i q}{n}\right) \right)^{\rm T}\, .
\end{split}
\end{equation}

For constant coefficients, the operator of shift in time from
$t_0$ to $t_1$ depends only on $t=t_1-t_0$:
$U(t_1,t_0)=U(t)=\exp Kt$. We can use
(\ref{eigenvectorscycleSYM}) and write
\begin{equation}
\begin{split}
&U(t)=\sum_{q=1}^n \exp(\lambda_q t) |r_q\rangle\langle l_q|\, , \\
&(U(t))_{js}=\sum_{q=1}^n \exp(\lambda_q t) r_{qj} l_{qs} \\
&\qquad \quad=\frac{1}{n}\sum_{q=1}^n \exp\left[t\left(\cos\frac{2\pi q}{n} -1\right)\right]
\cos\left((s-j)\frac{2\pi q}{n}+t \sin\frac{2\pi q}{n}\right) \, .
\end{split}
\end{equation}
This explicit formula allows us to compute all the necessary
quantities including the contraction coefficient
$\delta_{U(t)}$ (\ref{normSubSp}).

Now, let us produce the approximate formula for the same
symmetric system by mixers. First of all, let us represent the
solution for the cycle by the path summation formula.  With the
convention of cyclic enumeration, the set of paths
$\mathcal{I}_i$ started at $A_i$ is the sequence
\begin{equation}
\mathcal{I}_i=\left\{\begin{array}{l}
A_i \xrightarrow{k_i}  \, , \\
A_i \xrightarrow{k_i} A_{i+1} \xrightarrow{k_{i+1}}\, , \\
\cdots\cdots\cdots\cdots\cdots\cdots\cdots\cdots\\
A_i \xrightarrow{k_i} A_{i+1} \xrightarrow{k_{i+1}} A_{i+2} \xrightarrow{k_{i+2}}
\ldots \xrightarrow{k_{i+j-1}}A_{i+j+1}\xrightarrow{k_{i+j}} \, , \\
\cdots\cdots\cdots\cdots\cdots\cdots\cdots\cdots
\end{array}\right\}\, .
\end{equation}

\begin{figure}
\centering{
\includegraphics[width=0.5\textwidth]{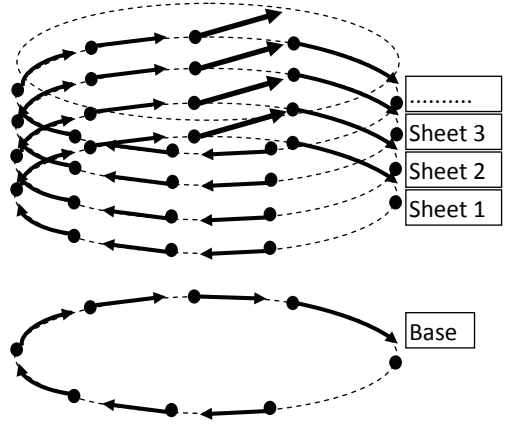}
}\caption{Multi--sheeted representation of the path summation
formula for a cycle (\ref{cycleGreen}): a cycle (the base) is
represented by an semi-infinite helix produced by redirecting of
reactions between sheets.
 \label{cycletower}}
\end{figure}

This sequence of paths corresponds to the multi--sheeted
representation presented in Fig.~\ref{cycletower}. First, we
consider a infinite series of the copies of the cycle. Each
vertex of the extended system is numerated by two indexes:
$(A_i,l)$, $i=1,2, \ldots , n$ (mod$n$), $l=1,2,3, \ldots$ is a
natural number. The reaction rate constants for copies are the
same as for the initial systems:
$k_{(j,r)(i,l)}=k_{ji}\delta_{rl}$. This extended system
obviously satisfies the definition of the multi--sheeted
extension of the cycle and in its projection on the base we
always have the kinetics of the cycle.

Let us select one number $i\in \{1, \ldots , n \}$ and recharge
the reactions: we annulate the ``horizontal" reaction rate
constant for $(A_i,l) \to (A_{i+1},l)$, $k_{(i+1,l)(i,l)}=0$,
and instead of this reaction take the reaction between levels,
$(A_i,l) \to (A_{i+1},l+1)$: $k_{(i+1,l+1)(i,l)}=k_{i+1\, i}$
(see Fig.~\ref{cycletower}). This is also a multi--sheeted
extension of the cycle. Formula (\ref{MSheetSum}) for this
multi--sheeted system allows us to use integration of the
infinite acyclic system (represented by the spiral in
Fig.~\ref{cycletower})) instead of integration of the finite
cyclic base system.

Now, let us put all $k_i=1$. For systems with constant
coefficients we use initial time moment $t_0=0$.  For the set
of paths $\mathcal{I}_i$ started at $A_i$ the solution to the
chain (\ref{ChainSimple}) with the initial conditions
$\varsigma_i(t_0)=1$ and $\varsigma_I=0$ for $|I|>1$ is
\begin{equation}\label{cyclePath}
\varsigma_I(t)=\frac{t^{|I|-1}}{(|I|-1)!}e^{-t}
\ .
\end{equation}
 Obviously, $\sum_{I\in \mathcal{I}_i} \varsigma_I =1$.
For concentration of $A_q$, formula (\ref{summation formula})
gives
\begin{equation}\label{cycleGreen}
u_{ji}(t)=e^{-t} \sum_{q=0}^{\infty}\frac{t^{qn+d_{ij}}}{(qn+d_{ij})!} \ ,
\end{equation}
where $d_{ij}$ is the length of the shortest oriented path from
$A_i$ to $A_j$ (here the length is the number of reactions and
the trivial path from $A_i$ to $A_i$ has the length zero).

For every two vertices $A_i$, $A_j$ we have only two mixers and
both are degenerated: $A_i  \xrightarrow{k} A_{i+1}
\xrightarrow{k} \ldots  \xrightarrow{k} A_j \xrightarrow{k}$,
length $j-i \mod n$ and $A_j  \xrightarrow{k} A_{j+1}
\xrightarrow{k} \ldots  \xrightarrow{k} A_i \xrightarrow{k}$,
length $i-j \mod n$.

Let us select one mixer $A_1  \xrightarrow{k} A_2 \ldots
\xrightarrow{k} A_j  \xrightarrow{k}$ for analysis. Initial
conditions are: $c_1=1$, $c_j=-1$ and other concentrations are
equal to zero.

For this auxiliary chain with given initial conditions
\begin{equation}\label{cycleMixer}
\begin{split}
&c_p=\frac{t^{p-1}}{(p-1)!}e^{-t} \ \ (p=1,\ldots , j-1), \\
&c_j=-e^{-t}\left(1-\frac{t^{j-1}}{(j-1)!}\right)
\ .
\end{split}
\end{equation}
The estimate (\ref{subsystemsEstimDeg}) $\|G^{ij}(t)\| \leq 1-
\int_{0}^{ t} {\Pi_S^+(\tau)} \ \D \tau$ is valid until $c_j$
changes its sign. Hence, for $t$ we have a boundary
$t^{j-1}\leq (j-1)!$. The Stirling formula gives a convenient
estimate:
\begin{equation}\label{Stirling}
\begin{split}
&t^{j-1}\leq \sqrt{2 \pi (j-1)}
\left(\frac{j-1}{e}\right)^{j-1} \lesssim (j-1)!\\
&t \leq t_1=\frac{j-1}{e}(2 \pi (j-1))^{\frac{1}{2(j-1)}} \ .
\end{split}
\end{equation}
Even a simpler estimate is $t<(j-1)/e$. If  $t$ satisfies one
of these inequalities then concentration $c_j$ is negative and
we can use the estimate (\ref{subsystemsEstimDeg}).

For this example,
\begin{equation}
\begin{split}\label{cycleestimeG}
&\Pi_S^+(t)=c_{j-1}(t)=\frac{t^{j-2}}{(j-2)!}e^{-t} \ , \;
\int_{0}^{ t} {\Pi_S^+(\tau)} \ \D
\tau=1-e^{-t}\sum_{p=0}^{j-2}\frac{t^p}{p!} \ , \\
&\|G^{ij}(t)\| \leq e^{-t}\sum_{p=0}^{\min\{d_{ji},d_{ij}\}-1}\frac{t^p}{p!} \, , \;
\delta_{U(t)} \leq e^{-t}\sum_{p=0}^{\left[\frac{n}{2}\right]}\frac{t^p}{p!} \, ,
\end{split}
\end{equation}
where $\left[\frac{n}{2}\right]$ is the integer part of $n/2$.
For $t>0$ this estimate gives $\|G^{ij}(t)\|<1$ and
$\delta_{U(t)}<1$ because $\sum_{p=0}^{j-2}\frac{t^p}{p!}<e^t$.
We can use the estimate (\ref{cycleestimeG}) on an interval
$[0,t_1]$, for example, on $[0,\frac{j-1}{e}]$. Intersection of
these intervals for all $i,j, i\neq j$ is $[0,\frac{1}{e}]$ ($j
\geq 2$). On this interval, the estimate (\ref{cycleestimeG})
is valid for all $i,j$. For extension of such an estimate for
$t>\frac{1}{e}$ the submultiplicative property (\ref{submulti})
can be used.

\section{Ergodicity Boundary and Limitation of Ergodicity}

In this Section we consider a reaction kinetic system
(\ref{kinurT}) with constant coefficients $k_{ji}>0$ for $(i,j)
\in \mathcal{E}$.

Let us sort the values of kinetic parameters in decreasing
order: $k_{(1)} > k_{(2)} > \ldots > k_{(n)}$. The number in
parenthesis is the number of value in this order. Each of the
constants $k_{(q)}$ is a reaction rate constant $k_{ij}$ for
some $i,j$ (and may be for several of them if values of these
constants coincide). Let us also suppose that the network is
weakly ergodic. We say that $k_{(r)}, \, 1 \leq r \leq n$ is
the
  {\em ergodicity boundary} \cite{GorOvRobu2007}
 if the network of  reactions with parameters $k_1,k_2, \ldots ,
k_r$ is weakly ergodic, but the network with parameters $k_1,k_2,
\ldots , k_{r-1}$ is not. In other words, when eliminating reactions
in decreasing order of their characteristic times, starting with the
slowest one, the ergodicity boundary is the constant of the first
reaction whose elimination breaks the ergodicity of the reaction
digraph.

Let $\mathcal{M}_{ij}$ ($i\neq j$) be a set of elementary
mixers (\ref{ElementaryMixer}), (\ref{ElementaryMixerDegen})
between given $A_i$, $A_j$. For each $M\in \mathcal{M}_{ij}$ we
can find a {\em cutting reaction rate constant}, ${\rm cut}_M$:
\begin{equation}
\begin{split}
{\rm cut}_M=\min\{k_{i_2i_1}, \ldots , k_{i_ri_{r-1}} , k_{i_r i_{r+1}}, \ldots ,
k_{i_{r+l-1}i_{r+l}}\} \ \ \mbox{for (\ref{ElementaryMixer})} \ ; \\
{\rm cut}_M=\min\{k_{i_2i_1}, \ldots , k_{i_ri_{r-1}}\} \ \ \mbox{for (\ref{ElementaryMixerDegen})} \ .
\end{split}
\end{equation}
Let us eliminate reactions in increasing orders of their constants (i.e. in decreasing order of their
characteristic times), starting with the smallest one. To cut all
elementary mixers between $A_i,A_j$ ($i\neq j$), it is necessary and sufficient to
eliminate all $k_{pq} \leq {\rm cut}_M$ for all $M \in \mathcal{M}_{ij}$. Therefore, for every
pair $A_i,A_j$ ($i\neq j$) we can also introduce a cutting constant:
$${\rm cut}_{ij}=\max_{M \in \mathcal{M}_{ij}}{\rm cut}_M \ .$$
To destroy the weak ergodicity of the network $\mathcal{N}$ we have to cut al least one pair
$A_i,A_j$ ($i\neq j$). The result can be formulates as the following theorem.

\begin{theorem}{Theorem 6} The ergodicity boundary of a network $\mathcal{N}$ is the following constant:
$${\rm cut}_\mathcal{N}=\min_{i\neq j} {\rm cut}_{ij} \ . \ \ \ \ \ \ \ \square$$
\end{theorem}

This boundary is a minimum (in pairs $A_i,A_j$) of maxima (in mixers $M \in \mathcal{M}_{ij}$) of
minima (in constants).

Kinetic equations for elementary mixers (\ref{ElementaryMixer}), (\ref{ElementaryMixerDegen})
allow explicit analytic solutions. Nevertheless, explicit estimates in terms of cutting constants
can be also useful.

Let for an elementary mixer $M$  (\ref{ElementaryMixer}) $\kappa_M$ be the maximal sum of constants
of outgoing reactions:

$$\kappa_M=\max\{\kappa_{i_p} \ | \ p=i_1,i_2, \ldots ,
i_{r+l} \}, \ \ \kappa_s= \sum_{p,\ p\neq s} k_{ps} \ ,$$

or for a degenerated elementary mixer $M$ (\ref{ElementaryMixerDegen})
$$\kappa_M=\max\{\kappa_{i_p} \ | \ p=i_1,i_2, \ldots , i_{r} \} \ .$$

Let us substitute all the constant for horizontal arrows in the elementary mixer
$M$ (\ref{ElementaryMixer}), (\ref{ElementaryMixerDegen})
by $k={\rm cut}_M$, and all the constants for vertical arrows ($i\neq i_r$)
by $\kappa-k$, where $\kappa=\kappa_M$. This change decreases the fluxes $\Pi^{\pm}$.

To find the estimate we have to solve the kinetic equation for a simple uniform kinetic path:
\begin{equation}\label{ElementaryMixerSimplif}
  \begin{CD}
A_{1} @>{k}>> A_{2} @>{k}>> \ldots @>{k}>> A_{s} @>{k}>>  \\
@VV{\kappa - k}V @VV{\kappa-k}V @. @VV{\kappa-k}V @. \\
  \end{CD}
\end{equation}
Similar to the simple cycle (\ref{cycleMixer}), we find
\begin{equation}\label{ElemMixerEstim}
c_p=\frac{(kt)^{p-1}}{(p-1)!}\exp(-\kappa t) \ \ (p=1,\ldots , s) \ ,
\end{equation}
the only difference is in exponents.

For the elementary mixers (\ref{ElementaryMixer}), (\ref{ElementaryMixerDegen}) this formula gives

$$\Pi^+(t) \geq k \frac{(kt)^{r-2}}{(r-2)!}\exp(-\kappa t) \ ,
\ \ \Pi^-(t) \geq k \frac{(kt)^{l-1}}{(l-1)!}\exp(-\kappa t) \
$$ and the estimates from Theorems~\ref{Theorem 4},\ref{Theorem 5} (\ref{subsystemsEstim}), (\ref{subsystemsEstimDeg})
become simple analytical expressions after substitution of $\Pi^{\pm}$ by their estimates from below.

Let us find an universal estimate from below for $t_1$. It is
$$\vartheta = \frac{1}{k+\kappa} \ .$$

Indeed, in the degenerated elementary mixer  (\ref{ElementaryMixerDegen})
on the way from $A_i$ to $A_j$ there exists at least one reaction with reaction rate constant
$k$: $A_r \to \ldots$. The integral flux through this reaction during the time interval $[0,t]$
is $$\int_0^t k c_r(\tau) \  \D \tau \geq \int_0^t \Pi^+(\tau) \ \D \tau \ .$$
The last inequality holds because all the flux in the mixer should go
through the reaction $A_r \to \ldots$ before it enters the last vertex.
On the other hand, $\int_0^t k c_r(\tau) \  \D \tau  \leq \int_0^t k \exp(-k \tau)  \  \D \tau$ (the last
integral corresponds to the case when all the concentration is collected at the initial moment
at $A_r$ and goes only through the reaction $A_r \to \ldots$).
Therefore,
$$\int_0^t \Pi^+(\tau) \ \D \tau  \leq 1 - \exp(-k\tau)\ .$$
From the condition (\ref{stopflux}) we find the estimate for
$t_1$ from below: $t_1 \geq \tau_1$, where $\tau_1$ is solution to
$$1 - \exp(-k\tau)=\exp(-\kappa \tau)\ . $$
We use convexity of exponential functions and substitute them
in this equation by linear approximation at point $\tau=0$:
$\exp(-x)>1-x$ ($x>0$); this gives us the estimate of $\tau_1$
from below: $\tau_1 < \vartheta=\frac{1}{k+\kappa}$.

For $t\in [0,\vartheta]$, $kt < 1$ and
$$1=\frac{(kt)^{0}}{0!} > \frac{(kt)^{1}}{1!}>\ldots
>\frac{(kt)^{r}}{r!}> \ldots  \ .$$
For each mixer $M$ we introduce the {\em length of mixer }
$d_M=\max\{r-2,l-1\}$ for (\ref{ElementaryMixer}) and $d_M=r-2$
for (\ref{ElementaryMixerDegen}). In these notations, each
mixer $M\in \mathcal{M}_{ij}$ gives the estimate: for $t\in
[0,\vartheta_M]$
\begin{equation}
\|G^{ij}(t)\| \leq 1 - \int_0^t {\rm cut}_M \frac{({\rm cut}_M \tau)^{d_M}}{(d_M)!}
\exp(-\kappa_M \tau) \ \D \tau \ ,
\end{equation}
where
$$\vartheta_M=\frac{1}{{\rm cut}_M + \kappa_M} \ .$$

For each pair $i,j$ ($i\neq j$) we can select the ``critical"
elementary mixer $M\in \mathcal{M}_{ij}$ with ${\rm cut}_M={\rm
cut}_{ij}$ and put $d_{ij}=d_M$, $\kappa_{ij}=\kappa_M$. If there
are several critical elementary mixers then we select one with
minimal $d_M$, if there are several such a mixers with minimal $d_M$
then we select one with minimal $\kappa_M$. In this notation we have
\begin{equation}
\|G^{ij}(t)\| \leq 1 - \int_0^t {\rm cut}_{ij} \frac{({\rm cut}_{ij} \tau)^{d_{ij}}}{(d_{ij})!}
\exp(-\kappa_{ij} \tau) \ \D \tau \
\end{equation}
for $t\in [0,\vartheta_{ij}]$, where
$$\vartheta_{ij}=\frac{1}{{\rm cut}_{ij} + \kappa_{ij}} \ .$$

Finally, for the whole network $\mathcal{N}$
$${\rm cut}_{\mathcal{N}}=\min_{i,j, i\neq j}\{{\rm cut}_{ij}\}, \
d_{\mathcal{N}}=\max_{i,j, i\neq j}\{d_{ij}\}, \ \kappa_{\mathcal{N}}=\max_{i,j, i\neq j}\{\kappa_{ij}\}, \
\vartheta_{\mathcal{N}}=\frac{1}{{\rm cut}_{\mathcal{N}} + \kappa_{\mathcal{N}}}$$
and for the contraction coefficient $\delta (t)$ (\ref{contractiondefinition}) we obtain the estimate
\begin{equation}\label{network contraction}
\begin{split}
\delta(t) \leq & 1 - \int_0^t {\rm cut}_{\mathcal{N}}
\frac{({\rm cut}_{\mathcal{N}} \tau)^{d_{\mathcal{N}}}}{(d_{\mathcal{N}})!}
\exp(-\kappa_{\mathcal{N}} \tau) \ \D \tau \\ &=1-
\left( \frac{{\rm cut}_{\mathcal{N}}}{\kappa_{\mathcal{N}}}\right)^{d_{\mathcal{N}}+1}
\left[1-\sum_{p=0}^{d_{\mathcal{N}}}
\frac{(\kappa_{\mathcal{N}} t)^p}{p!}\exp (-\kappa_{\mathcal{N}} t) \right]
\end{split}
\end{equation}
for $t\in [0,\vartheta_{\mathcal{N}}]$. For $t$ outside this interval, the submultiplicative
property (\ref{submulti}) should be used.

\section{Discussion}

The kinetic path summation formula together with the
multi--sheeted extension of kinetics provide us with a factory
of estimates. It is difficult to find, who invented this
approach.

The analysis of kinetic paths with selection of the most
important (dominant) paths allowed us to extract dominant
systems from kinetic equations
\cite{GorbaRadul2008,GorbanRadZinChemEngSci2010}. A robust
procedure for simplification of biochemical networks was
created \cite{RadGorZinLil2008}. This approach was developed
into unified framework for hybrid simplifications of Markov
models of multiscale stochastic gene networks dynamics
\cite{Debussche}. Dominant subsystems were analyzed for
dynamical models of microRNA action on the protein translation
process \cite{ZinatAlBMC2010}.

The multi--sheeted extension of kinetics provides us with a
simple and useful technique for estimation of relaxation
processes in Master equation. This method introduces an
internal ``microstructure" in the first order kinetic systems.
The kinetic path summation formula is a particular case of the
formula (\ref{MSheetSum}) (Proposition~\ref{Proposition 2}).

Indeed, let us construct the following multi--sheeted extension of
the Master equation. The set of components is $\mathcal{A} \times
\mathcal{K}$, where $\mathcal{K}=\{0\}\cup \mathcal{K}_1$ and
$\mathcal{K}_1$ is the set of all kinetic paths $I$ with lengths
$|I|>1$ (non-degenerated paths). The connections between sheets
(redirected reactions) are:
$$A_{i_{I^-},I^-} \xrightarrow{k_I} A_{i_I,I} \ \ \mbox{instead of}\ \
A_{i_{I^-},I^-} \xrightarrow{k_I} A_{i_I,I^-} \ . $$ According
to this rule, the reaction that continues the path $I^-$ to the
path $I$ is redirected and goes from the sheet $I^-$ to the
sheet $I$. For a degenerated $I^-$, we take
$A_{i_{I^-},I^-}=A_{i_{I^-},0}$, this means that all paths
start on the zero sheet, and all reactions from this sheet lead
to other sheets: $A_i \to A_j$ transforms into $A_{i,0} \to
A_{j,\{i,j\}}$, where $\{i,j\}$ is a path of the length 2.
Formula (\ref{MSheetSum}) for this multi--sheeted structure
coincide with the kinetic path summation formula
(\ref{summation formula}) (Theorem~\ref{Theorem 2}) for initial
conditions $c_{i,0}=1$ and other $c_{(j,I)}=0$.

This multi--sheeted extension may be considered as a generalization
of the Bethe lattices introduced by H. Bethe in 1935
\cite{Baxter1982}. For example, if in the initial graph of reactions
each vertex has the same number of outgoing edges then the
constructed multi--sheeted extension  can be considered as a bundle
of the Bethe lattices, each of them starts from one point of the
zeroth sheet. For each starting point, $A_{(i,0)}$ the corresponding
Bethe lattice represents the ``Green function" $u_{ji}(t,t_0)$ for
given $i$ and for all possible $j$.

We produced the kinetic path summation formula for
time--dependent kinetic equations and applied this formula for
evaluation of the ergodicity coefficient. The evaluation of the
contraction coefficient in the $l_1$ norm is the main tool for
studying of the relaxation in time--dependent Markov processes
since the seminal works of R. Dobrushin
\cite{DobrushinErgCoeff1956}.

Another important context of this study is the analysis of the
eigenvalues of the stochastic matrices
\cite{DmitrievDynkin1946,Karpelevich1951} and, especially the
analysis of these eigenvalues for matrices with specified graph
\cite{JohnKellog1_1978,JohnKellog2_1979}. In chemical kinetics,
evaluation of the eigenvalues through kinetic constants was
given in series of work by V.Cheresiz and G. Yablonskii
\cite{CherYab,YabCher}.

Various estimates of eigenvalues of $K$ could be produced from
the estimates of contraction (\ref{subsystemsEstim}), (\ref{subsystemsEstimDeg}).
The simplest one follows from (\ref{network contraction}):
\begin{equation}
Re(\lambda)\leq \frac{\ln(\delta(\vartheta))}{\vartheta} < 0 \ .
\end{equation}

Several problems should be resolved to make the use of the path
summation formula more effective. Perhaps, the most important
of them was mentioned in the comment \cite{FlomenbomSun2006}).
The amount of the kinetic path needed for accurate estimate of
the solution grows quickly in time for a sufficiently complex
system. Hence, we need either special tricks for the analysis
of path sampling or special asymptotic formulas for long paths
instead of exact solutions.

Another possible approach to this problem is in the use of more
complex exactly solvable systems instead of paths. The set of
reactions is solvable, if there exists a linear transformation of
coordinates $c \mapsto a$  such that kinetic equations in new
coordinates for all values of reaction constants have the triangle
form:
\begin{equation}\label{triangle}
\frac{\D a_i}{\D t}=f_i(a_1, a_2,... \, a_i).
\end{equation}
The algorithm for the analysis of reaction network solvability
was developed in \cite{Ocherki} (see also
\cite{GorbaRadul2008}). The simplest examples of solvable
networks give acyclic graphs (reaction trees) and pairs of
mutually inverse reactions. It may be possible to decompose the
complex system of transitions into a sequence of solvable
systems.

\end{document}